# Real-Time Character Inverse Kinematics using the Gauss-Seidel Iterative Approximation Method

Ben Kenwright
School of Computing Science
Newcastle University
Newcastle, United Kingdom
b.kenwright@ncl.ac.uk

*Abstract*—We present a realistic, robust, and computationally fast method of solving highly non-linear inverse kinematic problems with angular limits using the Gauss-Seidel iterative method. Our method is ideally suited towards character based interactive applications such as games. To achieve interactive simulation speeds, numerous acceleration techniques are employed, including spatial coherent starting approximations and projected angular clamping. The method has been tested on a continuous range of poses for animated articulated characters and successfully performed in all cases and produced good visual outcomes.

*Keywords-character animation; gauss-seidel; inverse kinematics; real-time*

## I.  INTRODUCTION

Character Inverse Kinematics (IK) is an important and challenging topic in the graphics and robotics community, and in employment by numerous applications in the film, animation, virtual reality, and game industry [1–6].

However, character-based models can be highly complex; even the most simplified models of 20-30 joints can generate a vast number of poses [5][6]. Whereby producing a simple pose to achieve a solitary task can produce ambiguous solutions that make the problem highly nonlinear and computationally expensive to solve. For example, even a straightforward task of reaching to pickup an object can be accomplished by means of any number of motions.

This paper focuses on how the Gauss-Seidel algorithm [7] can be employed to solve character IK problems; such as the biped shown in Figure 1. The Gauss-Seidel algorithm is an iterative, efficient, low memory method of solving linear systems of equations of the form $Ax=b$. Hence, we integrate the Gauss-Seidel iterative algorithm with a character IK problem to produce a flexible whole system IK solution for time critical systems such as games. This method is used as it offers a flexible, robust solution with the ability to trade accuracy for speed and give good visual outcomes.

Furthermore, to make the Gauss-Seidel method a practical IK solution for characters, it needs to enforce joint limits. We incorporate joint limits by modifying the update scheme to include an iterative projection technique. Additionally, to ensure *real-time* speeds we take advantage of spatial coherency between frames as a warm starting approximation for the solver. Another important advantage of the proposed method is the simplicity of the algorithm and how it can be easily configured for custom IK problems.

*The main contribution of the paper is the practical demonstration and discussion of using the Gauss-Seidel method for real-time character IK problem. Furthermore, we discuss constraint conditions, speedup approaches and robustness factors for solving highly non-linear IK problems in real-time.*

The rest of the paper is organized as follows. First, Section II gives a brief survey of related work. In Section III, we present the biped character model used for our simulations. Then, Section IV explains how the Jacobian matrix is calculated. Sections V to VIII primarily discusses the Gauss-Seidel algorithm and implementation details. Section IX presents the results. Finally, Section X draws conclusions and future work.

## II.  RELATED WORK

IK is a vital component that can be implemented using a wide range of solutions. We give a brief overview of existing, current, and cutting-edge approaches to help emphasis the different ways of approaching the problem; enabling the reader to see where our method sits.

In general, however, for very simple problems with just a few links, analytical methods are employed to solve the IK problem. Alternatively, for larger configurations, iterative numerical methods must be employed due to the complexity of the problem.

The character IK problem of finding solutions to poses that satisfies positional and orientation constraints has been well studied, e.g., [1], [6], [8], [9]. The problem is highly nonlinear, meaning there can be numerous solutions; hence, multiple poses fulfilling the constraint conditions. In practical situations, there can even be cases where no solution exists due to the poor placement of end-effectors. IK systems typically use cut down models, e.g., merely performing IK on individual limbs (as in body, arms, legs) [5], [10], [11]. This makes the problem computationally simpler and less ambiguous.

Numerous solutions from various fields of research have been implemented to solving the IK problem. The most popular method and the method upon which we base our iterative solution is the Jacobian matrix [6][12][13]. The Jacobian matrix method aims to find a linear approximation to the problem by modelling the end-effectors movements relative to the instantaneous systems changes of the links translations and orientations. Numerous different methods have been presented for calculating the Jacobian inverse, such as, Jacobian Transpose, Damped Least Squares (DLS), Damped Least Squares with Singular Value Decomposition (SVD-DLS), Selectively Damped Least Square (SDLS) [3], [4], [14–17].

An alternative method uses the Newton method; whereby the problem is formulated as a minimization problem from which configuration poses are sought. The method has the disadvantage of being complex, difficult to implement and computationally expensive to computer-per-iteration [13].

The Cyclic Coordinate Descent (CCD) is a popular real-time IK method used in the computer games industry [18]. Originally introduced by Wang et al. [19] and then later extended to include constraints by Welman et al. [1]. The CCD method was designed to handle serial chains and is thus difficult to extend to complex hierarchies. It has the advantage of not needing to formulate any matrices and has a lower computational cost for each joint per iteration. Its downside is that the character poses even with constraints can produce sporadic and unrealistic poses. However, further work has been done to extend CCD to work better with human based character hierarchies [2][5][20].

A novel method recently proposed was to use a Sequential Monte Carlo approach but was found to be computationally expensive and only applicable for offline processing [21][22].

Data driven IK systems have been presented; Grochow et al. [23] method searched a library of poses to determine an initial best guess solution to achieve real-time results. An offline mesh-based for human and non-human animations was achieved by learning the deformation space; generating new shapes while respecting the models constrains [24], [25].

A method known as 'Follow-The-Leader' (FTL) was presented by Brown et al. [26] and offered real-time results using a non-iterative technique. However, this approach was later built upon by Aristidou et al. [27] and presented an iterative version of the solver known as FABRIK.

The Triangular IK method [28][29], uses trigonometric properties of the cosine rule to calculate joint angles, beginning at the root and moving outwards towards the end-effectors. While the algorithm can be computationally fast, due to it being able to propagate the full hierarchy in a single iteration, it cannot handle multiple end-effectors well and is primarily based around singly linked systems.

The advantages of an iterative character IK system were also presented by a well written paper by Tang et al. [30] who explored IK techniques for animation using a method based on the SHAKE algorithm. The SHAKE algorithm is an iterative numerical integration scheme considered similar to the Verlet method [31], which can exploit substantial step-sizes to improve speed yet remain stable when solving large constrained systems. The algorithm is also proven to have the same local convergence criterion as the Gauss-Seidel method we present here as long as the displacement size is kept sufficiently small.

## III. ARTICULATED CHARACTER MODEL

We model the mechanical functioning of the biped as a series of multiple rigid segments (or links) connected by joints. This interconnected series is also called a kinematic chain.

As shown in Figure 1, we represent the biped character as a collection of 14 rigid body segments connected using 8 primary joints. The character gives us 30 degrees of freedom (DOF).

Joints such as the shoulder have three DOF corresponding to abduction/adduction, flexion/extension and internal/external rotation (i.e., rotation around the *x*, *y* and *z* axis).

Furthermore, it is convenient to note that a joint with *n* DOF is equivalent to *n* joints of 1 DOF connected by n-1 links of length zero. Thus, the shoulder joint can be described as a sequence of 3 separate joints of 1 DOF, where 2 of the joints connecting links have zero lengths.

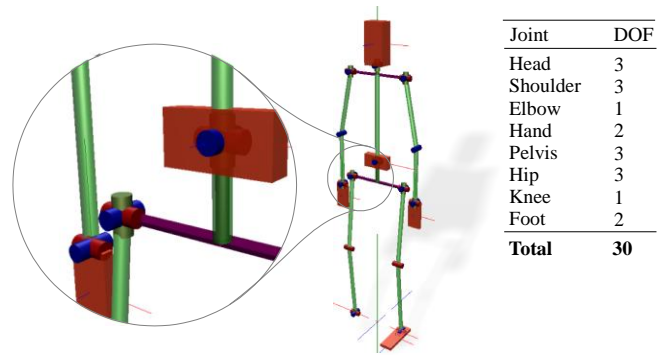

| Joint | DOF |
|---|---|
| Head | 3 |
| Shoulder | 3 |
| Elbow | 1 |
| Hand | 2 |
| Pelvis | 3 |
| Hip | 3 |
| Knee | 1 |
| Foot | 2 |
| **Total** | **30** |

Figure 1. The joint configuration with the right foot set as the IK base.

As shown in Figure 1, the single DOF connected joints were colored in accordance with their axis type; the *x*, *y* and *z* representing the colors red, green and blue. The foot was set as the base for the IK with five end-effectors (i.e., head, pelvis, right-hand, left-hand and left-foot). We developed an application for an artist to interrogate and experiment with the biped IK system; setting end-effectors locations and viewing the generated poses.

Each end-effector has a 6 DOF constraint applied to it; representing the target position and orientation. The ideal end-effectors are drawn in red, and the current end-effectors are drawn in green. This can be seen clearly in Figure 5, where the target end-effectors are located at unreachable goals.

## IV. JACOBIAN MATRIX

The Jacobian $J$ is a matrix that represents the change in joint angles $\Delta\theta$ to the displacement of end-effectors $\Delta e$.

Each frame we calculate the Jacobian matrix from the current angles and end-effectors. We assume a right-handed coordinate system.

To illustrate how we calculate the Jacobian for an articulated system, we consider the simple example shown in Figure 2. For a more detailed description see [3], [4], [14–17]. The example demonstrates how we decompose the problem and represent it as a matrix for a sole linked chain with a single three DOF end-effector. We then extend this method to multiple linked-chains with multiple end-effectors (each with six DOF) to represent the character hierarchy.

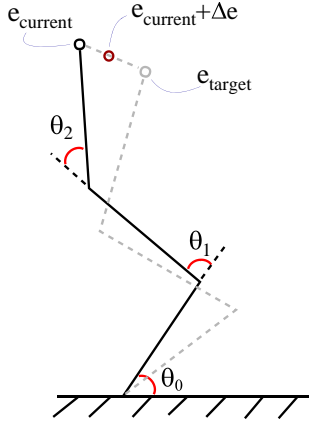

Figure 2. Relationship between multiple joint angles and end-effectors.

The angles for each joint and the error for each end-effector are represented by matrices.

$$\boldsymbol{\theta} = \begin{bmatrix} \theta_0 \\ \theta_1 \\ \theta_2 \\ ... \\ \theta_n \end{bmatrix} \quad (1)$$

$$\mathbf{e} = \begin{bmatrix} e_x \\ e_y \\ e_z \end{bmatrix} \quad (2)$$

where $\theta_i$ is the rotation of joint $i$ relative to joint $i-1$, and $e$ for the end-effectors global position.

From these matrices, we can determine that the end-effectors, and the joint angles are related. This leads to the forward kinematics definition, defined as:

$$\mathbf{e} = f(\boldsymbol{\theta}) \quad (3)$$

We can differentiate the kinematic equation for the relationship between end-effectors and angles. This relationship between change in angles and change in end-effectors location is represented by the Jacobian matrix.

$$\dot{\mathbf{e}} = \mathbf{J}\dot{\boldsymbol{\theta}} \quad (4)$$

The Jacobian $\mathbf{J}$ is the partial derivatives for the change in end-effectors locations by change in joint angles.

$$\mathbf{J} = \frac{\partial \mathbf{e}}{\partial \boldsymbol{\theta}} \quad (5)$$

If we can re-arrange the kinematic problem:

$$\boldsymbol{\theta} = f^{-1}(\mathbf{e}) \quad (6)$$

We can conclude a similar relationship for the Jacobian:

$$\dot{\boldsymbol{\theta}} = \mathbf{J}^{-1}\dot{\mathbf{e}} \quad (7)$$

For small changes, we can approximate the differentials by their equivalent deltas:

$$\Delta \mathbf{e} = \mathbf{e}_{target} - \mathbf{e}_{current} \quad (8)$$

For these small changes, we can then use the Jacobian to represent an approximate relationship between the changes of the end-effectors with the changes of the joint angles.

$$\Delta \boldsymbol{\theta} = \mathbf{J}^{-1}\Delta \mathbf{e} \quad (9)$$

We can substitute the result back in:

$$\boldsymbol{\theta}_{current} = \boldsymbol{\theta}_{previous} + \Delta \boldsymbol{\theta} \quad (10)$$

The practical method of calculating J in code is used:

$$\frac{\partial \mathbf{e}}{\partial \boldsymbol{\theta}_j} = \mathbf{r}_j \times (\mathbf{e}_{target} - \mathbf{p}_j) \quad (11)$$

where $\mathbf{r}_j$ is the axis of rotation for link $j$, $\mathbf{e}_{target}$ is the end-effectors target position, $\mathbf{p}_j$ is end position of link $j$.

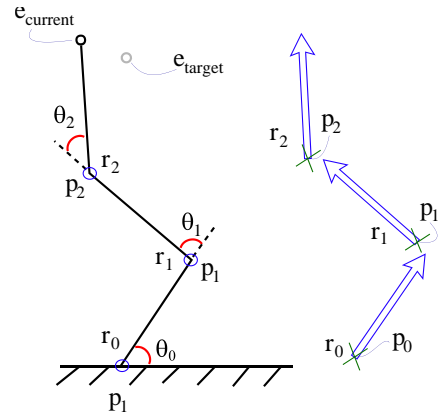

Figure 3. Iteratively calculating the Jacobian on a frame by frame basis.

For example, calculating the Jacobian for Figure 3 gives:

$$\mathbf{J} = \begin{bmatrix} \frac{\partial \mathbf{e}}{\partial \theta_0} \\ \frac{\partial \mathbf{e}}{\partial \theta_1} \\ \frac{\partial \mathbf{e}}{\partial \theta_2} \end{bmatrix} = \begin{bmatrix} \mathbf{r}_0 \times (\mathbf{e}_{current} - \mathbf{p}_0) \\ \mathbf{r}_1 \times (\mathbf{e}_{current} - \mathbf{p}_1) \\ \mathbf{r}_2 \times (\mathbf{e}_{current} - \mathbf{p}_2) \end{bmatrix} \quad (12)$$

and

$$\mathbf{e} = \mathbf{e}_{current} - \mathbf{e}_{target} \quad (13)$$

The Jacobian matrix is calculated for the system so that we can calculate the inverse and hence the solution.

Alternatively, a good explanation of the Jacobian and its applications is also presented by Buss [12], who gives an introduction to IK methods using the Transpose, Pseudoinverse, and Damped Least Square (DLS) method.

## V. FORMULATING THE GAUSS-SEIDEL PROBLEM

We set up the IK problem into a particular arrangement, so that we can solve for the unknowns using the Gauss-Seidel method. Whereby, we construct the IK formulation using the Jacobian matrix with the linear equation format of the form:

$$\mathbf{A}\mathbf{x} = \mathbf{b} \tag{14}$$

The IK problem is then composed as:

$$\mathbf{J}^T \mathbf{J} \Delta \theta = \mathbf{J}^T \Delta \mathbf{e} \tag{15}$$

Equating equivalent variables:

$$\begin{aligned} \mathbf{A} &= \mathbf{J}^T \mathbf{J} \\ \mathbf{b} &= \mathbf{J}^T \Delta \mathbf{e} \\ \mathbf{x} &= unknown \end{aligned} \tag{16}$$

With the Gauss-Seidel iterative method, we solve for the unknown $x$ value. To prevent singularities and make the final method more stable and robust we incorporated a damping value:

$$\mathbf{A} = (\mathbf{J}^T \mathbf{J} + \delta \mathbf{I}) \tag{17}$$

where $\delta$, is a small damping constant, typically *0.001,* and $\mathbf{I}$ is an identify matrix.

## VI. ITERATIVE GAUSS-SEIDEL IK SOLUTION

The Gauss-Seidel iterative algorithm is a technique developed for solving a set of linear equations of the form *Ax=b*. The method has gained a great deal of acclaim in the physics-based community for providing a computationally fast robust method for solving multiple constraint rigid body problems [32–34].

The iterative algorithm is based on matrix splitting [35], and its computational cost per iteration is O(n), where *n* is the number of constraints. Furthermore, the number of constraints and the number of iterations is what dominates the performance of the algorithm.

Algorithm 1 is the basic Gauss-Seidel method for a generic linear system of equations of the form *Ax=b*; for the unknowns, an initial guess $x^0$ is needed. Naively this value could be zero and result in the system having a cold start. Then the algorithm would proceed, while at each iteration, the corresponding elements from *A*, *b* and *x (current)* act as a feedback term to move *x (next)* closer to the solution.

The conditions for the algorithm terminating are:
- If a maximum number of iterations has been reached.
- If the error $\|Ax-b\|$ drops below a minimum threshold.
- If $\|\Delta x_i\|$ falls below a tolerance.
- If $\|\Delta x_i\|$ remains the same as the previous frame (within some tolerance).

$x = x^0$

**for** $iter = 1$ to iterationlimit **do**

    **for** $i = 1$ to $n$ **do**

$$\Delta x_i = \frac{\left[ b_i - \sum_{j=1}^{n} A_{ij} x_j \right]}{A_{ii}}$$

$$x_i = x_i + \Delta x_i$$

    **end for**

**end for**

Algorithm 1. Gauss-Seidel iterative algorithm to solve Ax=b given $x^0$.

It is essential that the coefficients along the diagonal part of the matrix be dominant for the Gauss-Seidel method to converge on a solution.

## VII. APPLYING JOINT LIMITS

Any IK solution needs to enforce angular joint limits before it can be a viable solution for a character system. We can modify the basic iterative algorithm to enforce joint limits by clamping the angles for each iteration update.

$$\theta = \begin{cases} lower & : if\ \theta + J^{-1}\Delta e < lower \\ upper & : if\ \theta + J^{-1}\Delta e > upper \\ \theta + J^{-1}\Delta e & : otherwise \end{cases} \tag{18}$$

This extension of the basic Gauss-Seidel algorithm to handle constraint limits for the unknowns is called the Projected Gauss-Seidel (PGS) algorithm. The angular limits form bounds that are in form of upper and lower joint angles that are easily enforced through clamping.

Furthermore, the PGS algorithm has O(n) running time and convergence is guaranteed as long as the matrix is positive definite [7]. In practice, we have found the algorithm to provide excellent visual and numerical results.

## VIII. SPATIAL AND TEMPORAL COHERENCY

To give the iterative solver an initial kick-start we take advantage of spatial and temporal coherency of the problem. Since the PGS solver is iterative by design the convergence rate can be slow; depending on the eigenvalues of the matrix. However, by caching the result from the previous solution, we can considerably reduce the number of iterations, especially if there are only minute changes.

## IX. RESULTS

### A. Walking

Setting the foot as the base of the IK solution and alternating it to the other foot between steps. At each foot changeover, a target trajectory was calculated for the swinging foot. The swinging foot end-effectors were interpolated to animate the walking motion. This produced the lower-body walking-motion. In addition, the upper-body end-effectors had similar trajectories calculated to induce upper-body motion to co-inside with the feet. As the trajectory interpolation between the start and end-effectors was performed, the IK solution was iteratively updated to generate a natural smooth blending animation with natural looking poses.

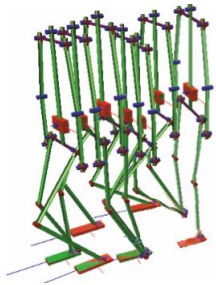

Figure 4. The step cycle for the IK; swapping the IK root between the left and right foot - with the standing foot holding the weight chosen as the root.

### B. Random Poses

We experimented with a diverse range of poses of generally unpredictable and chaotic stature to explore the stability and flexibility of our approach. For example, we did random on the spot poses of the character kneeling, standing on one single leg, waving, and so on (as shown in Figure 6).

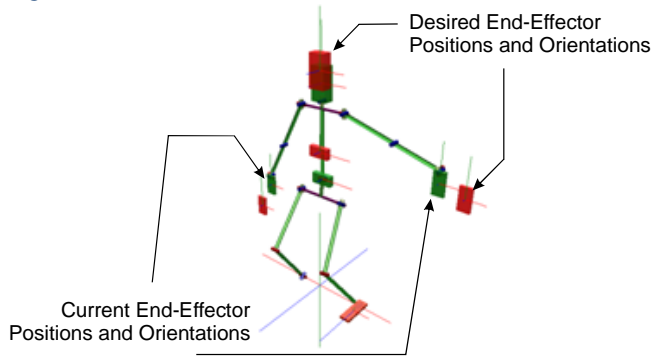

Figure 5. The Gauss-Seidel method remains stable even when we impose impossible constraint conditions on the end-effectors – the solver converges on achieving a best attempt stretching solution.

### C. Robustness

One important criteria was that the IK solver remained stable – this included placing end-effectors out of reach so that no solution could be found.

In practice, when no result was obtainable, a best reach condition was always presented, stretching to obtain the end-effectors but remaining stable (i.e., not oscillating or jittering).

Furthermore, when end-effectors were started at radically different locations, the resulting solution would radically jerk – however, the result always converged on acceptable poses.

### D. Performance

On average, the small spatial coherent transitions between frame updates resulted in the Gauss-Seidel method requiring only two or three iterations for the end-effectors to reach acceptable answers. This resulted in the IK solver being able to easily maintain a low-computational overhead and run at real-time frame-rates. Our Gauss-Seidel implementation was straightforward and single threaded; however, numerous methods have been demonstrated by Courtecuisse et al. [36] to exploit even greater performance improvements by taking advantage of multi-core architectures.

|  | Iterations | | |
| --- | --- | --- | --- |
|  | 1 | 5 | 20 |
| Avg. Time | 0.01ms | 0.042ms | 0.11ms |

Table 1. Performance of our Gauss-Seidel character implementation. Where little or no movement results in 1-2 iterations while sporadic changes in posture resulted in ~10 or more iterations.

Furthermore, our Gauss-Seidel method would only require a few mill-seconds to compute the solution. The cost of calculating the full IK biped solution for different iteration is shown in Table 1. Our implementation performed at real-time rates and maintained a consistent frame-rate well above a 100Hz.

Simulations were performed on a machine with the following specifications: Windows7 64-bit, 16Gb Memory, Intel i7-2600 3.4Ghz CPU. Compiled and tested with Visual Studio.

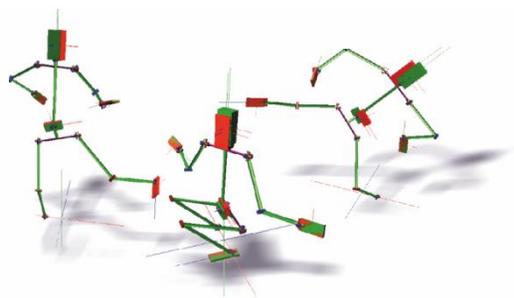

Figure 6. Examples of general disorderedly and chaotic investigation poses.

## X. CONCLUSION AND FURTHER WORK

We presented the Gauss-Seidel technique as a method for solving real-time character IK problems. We used temporal caching to reduce the computational cost and gain real-time performance speeds. The results of the IK system performed well enough to be used in time critical systems (such as games.) With the angular limits, the method can suffer from singularity problems if the end-effectors jump; however, due to the end-effectors following small spatial transitions singularities are mostly avoided.

The algorithm is simple to implement, computationally fast, little memory overhead, and is fairly robust. The IK solution can work with multiple end-effectors to produce poses with smooth movement with and without constraints.

While we demonstrated the practical aspect of using the Gauss-Seidel method as a valid real-time method for a character IK system, further work still needs to be done for a more detailed statistical comparison between the aforementioned IK solutions; comparing memory, complexity and computational costs.

In additional, a further area of study would be the general practical applicability of generating primary and secondary IK goals using weighted biasing of conflicting constraints (e.g., secondary goal of keeping body within balancing stance while always achieving the primary target of arms and feet reaching their goals.)